# Inverse Transverse Magneto-Optical Kerr Effect

V.I. Belotelov[1,2], A.K. Zvezdin[1]

[1]A.M. Prokhorov General Physics Institute RAS, 38 Vavilov st., Moscow, 119991 Russia, [2]M.V. Lomonosov Moscow State Univ., Moscow, 119991, Russia

**Abstract** It is demonstrated that a static in-plane magnetic field is generated in a ferromagnetic film by *p*-polarised light obliquely incident on the film. This phenomenon can be called inverse transverse magneto-optical Kerr effect. The femtosecond laser pulse of peak intensity of 500 W/μm² generates in nickel an effective magnetic field of about 100 Oe. The value of the effective magnetic field can be increased by more than an order of magnitude at the surface plasmon polariton resonance excited in smooth metal dielectric structures or in plasmonic crystals.

Exciting possibility of ultrafast control of a medium magnetization at subpicosecond time scale via light was demonstrated recently [1-3]. It is of prime importance for modern magnetic storage systems demanding very large operation rates. In particular, optical way of magnetization control is possible due to the fact that a circularly polarized light induces static magnetization in a gyrotropic medium. This phenomenon is called inverse Faraday effect. It was theoretically predicted by Pitaevskii [4] and Pershan [5] about 50 years ago. Its experimental evidence in paramagnets was delivered in [6]. The inverse Faraday effect is related to its well known counterpart – Faraday effect, or magnetic circular birefringence, resulting in rotation of the light polarisation through a magnetic medium. In the approximation of non-absorbing media they both are proportional to the same magneto-optical constant [7].

It is not the only Faraday effect which was shown to have its inverse counterpart. Thus, the inverse Cotton-Mouton effect was observed recently in terbium gallium garnet crystal [8]. Due to this effect, the medium magnetization is induced in a medium by a linearly polarized light propagating in the presence of a transverse magnetic field.

Apart from Faraday and Cotton-Mouton effects there are effects belonging to a family of magneto-optical Kerr effects. Among them transverse magneto-optical Kerr effect (TMOKE) is of prime importance providing an efficient tool for magnetization monitoring in different types of magnetic samples [9-10]. Furthermore, TMOKE can be used for optical reading in magnetic storage systems based on media with in-plane magnetization.

TMOKE is observed for oblique incidence of *p*-polarized electromagnetic wave in the plane perpendicular to the sample magnetization [11]. It is measured by a relative change of the sample reflectivity when the sample is remagnetized. Thus TMOKE belongs to the class of the intensity magneto-optical effects in the contrary to the Faraday or polar Kerr effects which are related to the light polarisation rotation. The origins of the effect are associated with the magnetization sensitive boundary conditions at the surface of the magnetic layer. The TMOKE takes place only for absorbing media.

In analogy to the case of the Faraday effect a question arises: whether the inverse counterpart of TMOKE exists. In the present Letter we address this problem and demonstrate by rigorous solution of Maxwell equations that this question has positive answer. We show that the static effective magnetic field is generated in a metal ferromagnet if it is illuminated by light in TMOKE configuration. The strength of the generated magnetic field is directly proportional to the incident intensity and is about 120 Oe for the incident light intensity of 500 W/μm². It should be noted that laser pulse of similar intensity ($10^3$ W/μm²) was used in the recent experiments on the inverse Faraday effect [2]. The inverse TMOKE can be significantly enhanced up to 5000 Oe by using plasmonic structures.

Within the macroscopic theory of magnetooptical phenomena the properties of the magnetically ordered media are defined by $\hat{\varepsilon}$- and $\hat{\mu}$-tensors. However, at the optical and near infrared frequencies a magnetic dipole response is very weak and $\hat{\mu}$-tensor is close to a scalar form and can be taken to be unity [7]. Magneto-optical properties of the medium are mainly governed by the gyration vector **g** which is a function of the medium magnetization **M**. For ferromagnetic crystals $g_i = a_{ij} M_j$, where tensor $a_{ij}$ is defined by the crystallographic symmetry. In the case of a medium with absorption gyration vector and tensor $a_{ij}$ become a complex valued function of frequency: $\mathbf{g} = \mathbf{g}' + i\mathbf{g}''$, $a_{ij} = a'_{ij} + ia''_{ij}$. In the linear in magnetization approximation the $\hat{\varepsilon}$-tensor of a ferromagnet is given by

$$\varepsilon_{ij} = \varepsilon_{ij}^0 - ie_{ijk} g_k, \qquad (1)$$

where $e_{ijk}$ is the Levi-Civita tensor.

Considering gyrotropic part of the energy of a ferromagnetic sample illuminated by light in the form of $U_M = \frac{i}{16\pi} e_{ijk} g'_k E_i^* E_j$ leads to the following expression for $\mathbf{H}_{eff}$:

$$H_i^{eff} = -\frac{i a'_{ij}}{16\pi} e_{jkl} E_k E_l^*. \qquad (2)$$

For the case of a ferromagnet crystal with cubic symmetry (2) simplifies to

$$\mathbf{H}^{eff} = -\frac{i a'}{16\pi} \left[ \mathbf{E} \times \mathbf{E}^* \right]. \qquad (3)$$

This formula is similar to the one for light induced magnetization of paramagnets [4]. The cross product of $\left[ \mathbf{E} \times \mathbf{E}^* \right]$ is not zero for elliptically polarized light.

However, existence of non zero $\left[\mathbf{E}\times\mathbf{E}^*\right]$ does not always demand the elliptical polarization of the incident light. Indeed, the cross product does not vanish if there is a phase shift between any of two orthogonal components of the electric field of the light wave inside a medium. This condition can be satisfied even for linear polarized light. In particular, it is possible to search for the inverse TMOKE described by light induced effective magnetic field in the TMOKE configuration with illumination by *p*-polarized light (electric field lies in the plane of incidence). It should be noted that *s*-polarized light (electric field is perpendicular to the incidence plane) has only one component of the electric field in a medium and consequently induces no effective magnetic field (provided that weak magnetic-dipole transitions are neglected). If electromagnetic wave with electric field $E^i$ is obliquely incident in the *xz*-plane, the angle of incidence is $\varphi$ and the magnetization vector **M** is along *y*-axis (Fig.1) then the electric field of the electromagnetic wave inside the ferromagnet is given by

$$E_x = -E^t \cos\varphi_t \exp(-i\gamma z),$$
$$E_z = -E^t (\sin\varphi_t + i\, g/\varepsilon \cos\varphi_t)\exp(-i\gamma z),\quad (4)$$

where $E^t = t_{12}E^i$, $t_{12} = 2\cos\varphi_i/(\tilde{n}\cos\varphi_i + \cos\varphi_t)$, $\varphi_t$ is the angle defined by Snell's law $\sin\varphi_t = \tilde{n}^{-1}\sin\varphi$ with complex refractive index of the ferromagnet $\tilde{n} = n(1+i\kappa)$, and $\gamma = k_0\tilde{n}\cos\varphi_t$. The angle $\varphi_t$ is generally complex valued. It is convenient to take $\cos\varphi_t$ in the form:
$\cos\varphi_t = q\exp(i\delta)$, with $q = (\alpha^2 + \beta^2)^{1/4}$,
$\alpha = 1 - (1-\kappa^2)n^{-2}(1+\kappa^2)^{-2}\sin^2\varphi_i$,
$\beta = 2\kappa n^{-2}(1+\kappa^2)^{-2}\sin^2\varphi_i$, $\tan 2\delta = \beta/\alpha$.

It follows from (3) that in the considered case the effective magnetic field is directed along *y*-axis and is written as

$$H_{eff} = \frac{ia'|E_i|^2|t_{12}|^2\exp(2\gamma''z)}{16\pi}\left[\sin\varphi_t(\cos\varphi_t)^* - (\sin\varphi_t)^*\cos\varphi_t\right], (5)$$

where $\gamma''$ is the imaginary part of $\gamma$. Here we assumed that the ferromagnet has cubic crystal symmetry and retrieved only terms linear in magnetization (linear in *a*). Usually, the refractive index of the ferromagnets has rather large real and imaginary parts and (5) can be simplified taking $\cos\varphi_t \approx 1$:

$$\mathbf{H}_{eff} = \frac{a'|E_i|^2|t_{12}|^2 \kappa\exp(4k_0 n\kappa z)}{8\pi n(1+\kappa^2)k_0}\left[\mathbf{k}_0\times\mathbf{N}\right]. \quad (6)$$

Direction of $\mathbf{H}_{eff}$, is determined by the cross product of the incident wave wavevector $\mathbf{k}_0$ and normal to the ferromagnet surface **N**. Consequently, $\mathbf{H}_{eff}$ reverses for the incident angle change from $\varphi_i$ to $-\varphi_i$.

It is vivid that for the considered configuration the effective magnetic field appears only for the materials having imaginary part of refractive index. It means that the electromagnetic wave has to decay inside the ferromagnetic medium which is the case if the material exhibits optical losses or/and has metallic nature with negative real part of the dielectric permittivity. However, if the magnetic film is sufficiently thin so that light reaches its bottom face and reflects back then the effective magnetic field does not vanish even for the medium with purely real refractive index.

There is an important difference between inverse TMOKE and well-known inverse Faraday effect. The inverse Faraday effect field $\mathbf{H}_{eff}$ is induced by circular polarized light and is directed along the wave vector $\mathbf{k}$. On the contrary, in accordance to (6) $\mathbf{H}_{eff}$ generated in the inverse TMOKE is oriented along the cross product of $\left[\mathbf{k}_0\times\mathbf{N}\right]$, i.e. $\mathbf{H}_{eff}$ is perpendicular to $\mathbf{k}_0$. The latter broadens experimental possibilities for investigation of the optically induced femtosecond magnetism. Note that the difference of these effects follows from the fact that corresponding linear magneto-optical effects have different dependence on relative orientations of the wavevector and magnetization **M** (or an external magnetic field). The Faraday angle $\psi \sim (\mathbf{k}\cdot\mathbf{M})$, but in TMOKE $\Delta R/R \sim (\mathbf{k}_0\cdot[\mathbf{M}\times\mathbf{N}])$, where $\Delta R/R$ is relative change of reflection.

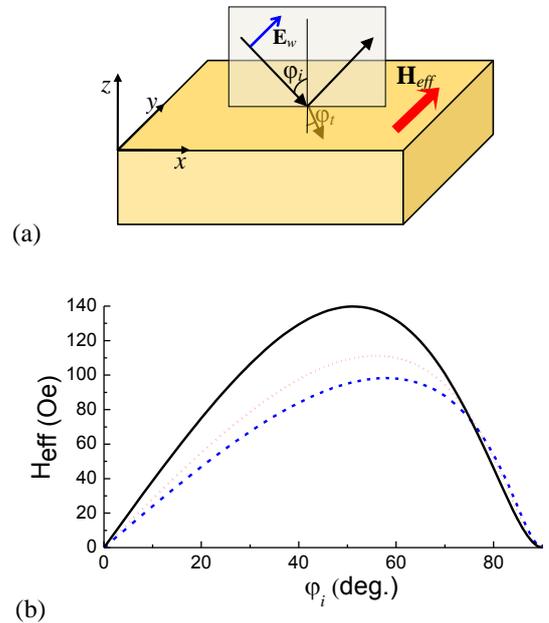

(a)

(b)

FIG.1. (a) Configuration for the inverse TMOKE: *p*-polarised light is obliquely incident on the ferromagnet film and generates an effective magnetic field $H_{eff}$ inside the film. (b) The inverse TMOKE near the surface (5 nm in depth) of magnetic films of iron (black solid line) at λ=630 nm, nickel (red dotted line) at λ=630 nm, and bismuth iron-garnet (blue dashed line) at λ=400 nm versus angle of light incidence. It is assumed that the samples are illuminated with a 40 fs duration laser pulse of peak intensity of 500 W/μm².

The inverse TMOKE is also not related to the magnetic dichroism. Indeed, the magnetic dichroism is determined by the imaginary part of the medium gyrotropy constant whereas it follows from (2) that the inverse TMOKE is related to the real part of the gyrotropy and thus exists even in the medium without any absorption and dichroism. The inverse TMOKE is also quite different from the inverse Cotton-Mouton effect since the latter being T-even effect requires the presence of an external magnetic field to produce a magnetic torque. On the contrary the inverse TMOKE can be observed without any external magnetic field.

The effective magnetic field generated by the electromagnetic field pulse in iron, nickel and bismuth iron-garnet films is shown in Fig. 1b. The effect acquires its maximum values at the oblique incidence of about 60 deg. At the maxima $H_{eff}$ is rather strong and is about 120 Oe near the film surface. For normal incidence the effect vanishes due to the absence of the $E_z$ component in the excitation electromagnetic field. Gyration of the dielectric iron-garnets is usually much smaller than the gyration of metallic ferromagnets. However near the absorption resonance at wavelength of about $\lambda=370$ nm the gyration can be rather large ($g \sim 0.1$) and $H_{eff}$ is compared to the one for the ferromagnetic metals (see blue dashed curve in Fig.1b).

The generated magnetic field exists only during the illumination of the sample and can be identified by its influence on the sample's magnetization. Namely, it can lead to magnetization precession around the external magnetic field. Magnetization dynamics can be observed experimentally in the pump-probe technique by measuring TMOKE and Faraday effect in the probe beam. The other consequence of the inverse TMOKE which can be observed in experiment is the appearance of electric current pulse in the ferromagnet in the direction along the light incidence (x-axis in Fig.1). Its value is estimated to be about 50 – 100 mA.

Recent progress in magnetoplasmonics demonstrated that most of the magneto-optical effects can be significantly enhanced due to excitation of surface plasmon polaritons (SPP) [12-16]. Moreover, SPP assisted magnetization of a nanohole array in noble metals was demonstrated recently by excitation of SPPs by circularly polarized light [17]. The most gain coming from the SPP generation takes place in structures with relatively low optical losses. One of the examples of such structures is a system of a structured noble metal and a magnetic dielectric operating at light frequency far from the electronic transitions [12,13]. That is why the enhancement factor for plasmonic structures with ferromagnet metals like nickel or cobalt is rather small [14]. Nevertheless, if the structure incorporates a combination of a noble and a ferromagnetic metals the SPP significance can be partly rescued [15,16]. We in what follows consider the inverse TMOKE in several plasmonic structures.

In the simplest case the SPP can be excited on a smooth ferromagnetic metal surface using prism in Kretschmann or Otto configurations (Fig.2) [18]. It provides the increase of $H_{eff}$ by about 4 times with respect to the non-plasmonic case. Further enhancement of the inverse TMOKE can be achieved if a periodic composite of alternating gold and nickel subwavelength stripes is considered (see inset to Fig.2). For the wavelength of 630 nm gold and nickel permittivities have close real parts but their imaginary parts differ much ($\varepsilon_{Au}=-8.9+1.1i$ and $\varepsilon_{Ni}=-9.2+14.4i$ [19]). So wavelengths of the SPPs waves at the gold and nickel surfaces are almost the same but SPPs absorption coefficients are much different. The wavelength of SPP for the gold/nickel composite is 595 nm (at $\lambda=630$ nm). If the period of the structure is much less compared to this value then the gold/nickel structure acts as a kind of an effective medium. The smaller filling factor of nickel in the composite the larger efficiency of the SPP excitation.

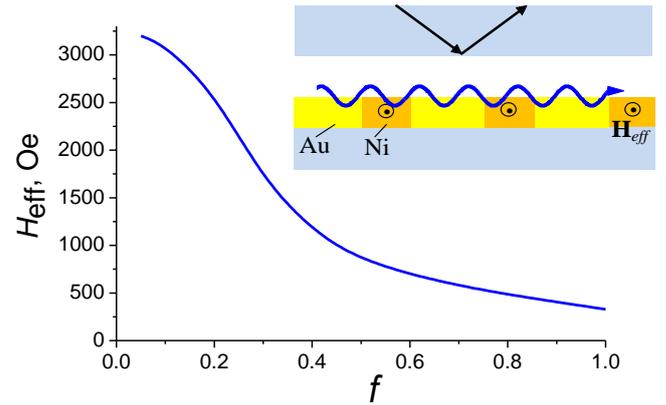

FIG.2. The inverse TMOKE at the SPP resonance in a Au/Ni periodic composite (shown in the inset) versus filling factor of the nickel part $f$ ($\lambda=630$ nm). Effective magnetic field $H_{eff}$ is calculated 5 nm in depth of the composite. Composite period $d$ is 200 nm, thickness is 200nm and substrate is a dielectric of $\varepsilon=5$ (blue lower region). The SPP is excited in Otto configuration via prism of $\varepsilon=5$ (blue upper region) by $p$-polarised laser pulse incident at 28°. Pulse peak intensity is 500 W/$\mu$m$^2$. Pulse duration is 40 fs.

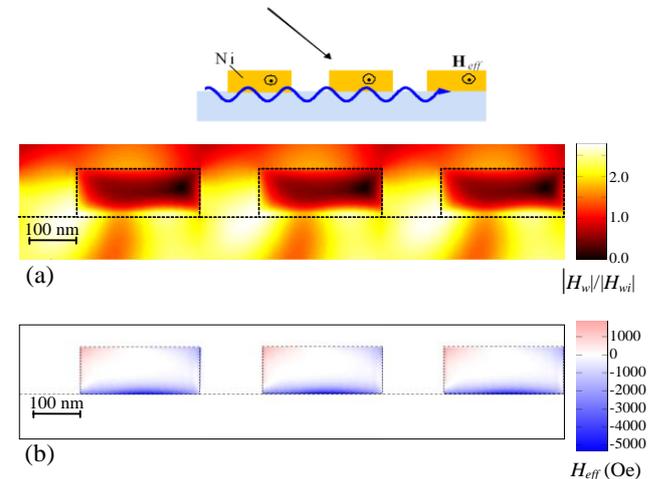

FIG.3. (a) Near field distribution of the absolute value of the electromagnetic wave magnetic field $|H_w|$ in the plasmonic crystal (shown in the inset above) normalized to the magnetic field of the incident light $|H_{wi}|$. (b) Distribution of the effective magnetic field $H_{eff}$ generated in the plasmonic crystal by the laser pulse. Three periods of the structure are shown. Laser pulse of peak intensity of 500 W/$\mu$m$^2$ is $p$-polarized and is obliquely incident at 10 deg. Period $d$ is 400 nm, slit width is 120 nm, and nickel film thickness is 100 nm.

Uniform nickel film corresponds to the filling factor $f=1$, where $f$ determines relative volumetric fraction of nickel in the composite. If $f$ gets smaller then $H_{eff}$ increases which is due to the enhancement of the SPP assisted electromagnetic field concentration near the composite interface. The value of $H_{eff}$ reaches 3000 Oe for very thin nickel stripes ($f < 0.1$). Consequently, the SPP introduces a significant increase of the inverse TMOKE by more than an order of magnitude.

If gold stripes are eliminated from the considered composite and period of the composite is made comparable to the SPP wavelength then the effects of the structure periodicity prevail and the structure can be referred to as a plasmonic crystal. Plasmonic crystals of perforated gold and smooth iron-garnet were shown recently to provide significant enhancement of TMOKE [12]. Though nickel plasmonic crystal has much larger optical losses it can also concentrate electromagnetic energy at the SPP resonance conditions (Fig. 3a). The electromagnetic energy density near the nickel/silica increases by an order of magnitude with respect to the one of the incident wave.

Generated effective magnetic field in the plasmonic crystal also increases and exceeds 5000 Oe near the nickel interface which corresponds to almost two fold further increase of $H_{eff}$. Prominent feature of the inverse TMOKE in plasmonic crystals is that $H_{eff}$ has opposite direction in different parts of the nickel stripe crossection (shown with blue and pink colors in Fig. 3b). It opens a new opportunity for the magnetization control of the ferromagnetic materials.

To conclude we have shown analytically and numerically that if a magnetically ordered material is obliquely illuminated with *p*-polarised light then inverse TMOKE takes place and static effective magnetic field is generated. It is directed perpendicularly to the incidence plane. If the laser pulse is of high intensity (of about 500 W/μm$^2$) $H_{eff}$ in nickel is about 100 Oe (at λ=630 nm). The effective magnetic field exists only during the laser pulse propagation inside the medium. It can be detected via observation of the magnetization dynamics caused by the inverse TMOKE or by measuring electric current pulse. Since $H_{eff}$ is determined by the near field electromagnetic field distribution the inverse TMOKE can be substantially enhanced at the SPP resonances providing significant electromagnetic energy concentration. Large values of the effective magnetic field generated in the inverse TMOKE can allow efficient all-optical magnetization control of magnetically ordered materials.

Work is supported by Russian Foundation of Fundamental Research, Russian Federation Presidential Grant No. MK-3123.2011.2 and by the Ministry of Education and Science of the Russian Federation (No. 16.740.11.0577).

REFERENCES

[1] A. V. Kimel, A. Kirilyuk, P. A. Usachev, R. V. Pisarev, A. M. Balbashov, and Th. Rasing, Nature **435**, 655 (2005).
[2] A. Kirilyuk, A.V. Kimel, and Th. Rasing, Rev. Mod. Phys. **82**, 2731 (2010).
[3] K. Vahaplar, A. M. Kalashnikova, A. V. Kimel, D. Hinzke, U. Nowak, R. Chantrell, A. Tsukamoto, A. Ithoh, A. Kirilyuk, and Th. Rasing, Phys. Rev. Lett. **103**, 117201 (2009).
[4] L. P. Pitaevskii, Sov. Phys. JETP **12(5)**, 1008 (1961).
[5] P. S. Pershan, Phys. Rev. **130**, 919 (1963).
[6] J. P. van der Ziel, P. S. Pershan, and L. D. Malmstrom, Phys. Rev. Lett. **15(5)**, 190 (1965).
[7] L. D. Landau, and E. M. Lifshitz, *Electrodynamics of Continuous Media.* (Pergamon Press, 1984).
[8] A. Ben-Amar Baranga, R. Battesti, M. Fouché, C. Rizzo, and G. L. J. A. Rikken, Eur. Phys. Lett. **94,** 44005 (2011).
[9] A. Chizhik, A. Zhukov, J.M. Blanco, and J. Gonzalez, JMMM **249**, 27 (2002).
[10] A. A. Rzhevsky, B. B. Krichevtsov, D. E. Bürgler, and C. M. Schneider, Phys. Rev. B **75**, 144416 (2007).
[11] A. K. Zvezdin, V. A. Kotov. *Modern Magnetooptics and Magnetooptical Materials* (IOP, Bristol, 1997).
[12] V. I. Belotelov, I. A. Akimov, M. Pohl, V. A. Kotov, S. Kasture, A. S. Vengurlekar, A. V. Gopal, D. Yakovlev, A. K. Zvezdin, and M. Bayer, Nature Nanotechnolgy **6**, 370 (2011).
[13] V.I. Belotelov, A.K. Zvezdin, JMMM **300**, e260 (2006).
[14] J. B. Gonzalez-Diaz, A. García-Martín, G. Armelles, J. M. García-Martín, C. Clavero, A. Cebollada, R. A. Lukaszew, J. R. Skuza, D. P. Kumah, and R. Clarke, Phys. Rev. B **76**, 153402 (2007).
[15] V. Temnov, G. Armelles, U. Woggon, D. Guzatov, A. Cebollada, A. Garcia-Martin, J. M. Garcia-Martin, T. Thomay, A. Leitenstorfer, and R. Bratschitsch, Nature Photonics **4**, 107 (2010).
[16] V.I. Belotelov, E.A. Bezus, L.L. Doskolovich, A.N. Kalish, A.K. Zvezdin, J. Phys: Conf. Ser. **200**, 092003 (2010).
[17] I.I. Smolyaninov, C.C. Davis, V.N. Smolyaninova, D. Schaefer, J. Elliott, and A.V. Zayats, Phys. Rev. B **71**, 035425 (2005).
[18] H. Raether, *Surface Plasmons on Smooth and Rough Surfaces and on Gratings*, (Springer-Verlag, New York, 1988).
[19] M. A. Ordal, R. J. Bell, R. W. Alexander, L. L. Long, and M. R. Querry, Appl. Opt. **24**, 4493 (1985).